\documentclass[aps,amsfonts,twocolumn,pra,superscriptaddress,10pts]{revtex4}
\usepackage[utf8]{inputenc}
\usepackage{graphicx}
\usepackage{subfig}
\begin{document}
\newcommand{\bra}[1]{\left\langle#1\right|}
\newcommand{\ket}[1]{\left|#1\right\rangle}
\newcommand{\abs}[1]{\left|#1\right|}
\newcommand{\mean}[1]{\left\langle #1\right\rangle}
\newcommand{\braket}[2]{\left\langle{#1}|{#2}\right\rangle}, 
\newcommand{\commt}[2]{\left[{#1},{#2}\right]}
\newcommand{\tr}[1]{\mbox{Tr}{#1}}

\title{Reviewing pigeonhole effect with transversal modes}
\author{Marcin Wie\'sniak}\affiliation{Institute of Theoretical Physics and Astrophysics, Faculty of Mathematics, Physics, and Informatics,\\ University of Gda\'nsk, ul. Wita Stwosza 57, 80-308 Gda\'nsk, Poland}
\affiliation{International Centre for Theory of Quantum Techologies\\ University of Gda\'nsk, ul. Wita Stwosza 63, 80-308 Gda\'nsk, Poland}
\begin{abstract}
We investigate the interference setup claimed to reveal the quantum pigeonhole effect. It is a claim that it is possible that three parties travel across a two-path interferometer, but no pair of them share a path. We demonstrate the inconsistency of the description based on the weak values.  We introduce transversal modes and observe the arise of correlations between path and traversal degrees of freedom. Also, the rotational symmetry of the output pattern is studied as a witness of pairwise interaction. It shows no apparent relation to the probability of observing three particles in the same output.  
\end{abstract}
\maketitle
\section{Introduction}
Intuitively we expect that when we send, say, three particles from one place to the other, each of should take some path from the initial to the final location. If we restrict possible paths to two possibilities, we expect that at least two them propagate together. Consequently, if these particles repulse one another, we would observe the witness of this repulsion in terms of a change of the observed output pattern.

These beliefs are challenged so-called quantum pigeonhole effect \cite{pigeon}. The claim of the original work is that, in a specific situation, three electrons travel through a two-path interferometer, but, basing on the weak value description, no pair of particles ever travelled together. This claim was subsequently generalized for $N$ particles with at most $M$ traveling through the same arm \cite{FOOTPRINTS}. In this work we reexamine these claims by introducing spatial degrees of freedom.

Quantum pigeonhole effect arose as one of the attempts of understanding interference. Traditional quantum mechanics is extremally successful in predicting results of experiments, but it forbids the observer to trace a particle inside an interferometer, leaving the observer with the nonclassical concept of superposition. Any attempt to localize a particle inside an inteferometer must result in the decrease of the interferometric visibility \cite{BERGE}. An alternative approach includes Bohmian mechanics and the two-state vector formalism (TSVF) and introduces some description inside an interferometer, but at the cost of many extra entities, such as nonlocal potentials, weak traces of presence, etc., and some paradoxes and counterintuitive claims, such as discountinuous trajectories \cite{Danan}. The core of TSVF is the realistic interpretation of weak measurements.

We begin with recalling the original presentation of the quantum pigeohole effect. We point out the inconsistency of the realistic interpretation of weak values, and stress the importance of the Hamming distance. Subsequently, we compare a quantity that witnesses propagation of two particles in the same arm and the probability of observing all three parties at the same side of the second beam-splitter. This probability is in a way complementary to the model of propagation expected in Ref. \cite{pigeon}. We find no clear relation between these two quantities.

\section{Quantum pigenhole effect}
The effect supposedly appears when we send three interacting particles 	(electrons) are sent through a suitably tuned (for the purpose of demonstrating the effect) two-path Mach-Zehnder interferometer, with arms denoted as $L$ and $R$. We assume that the interaction is only possible when two electrons are in the same arm. The state at the input is $\ket{LLL}$. The first beam splitter transforms it to $\ket{+++}$, where $\ket{+}=1/\sqrt{2}(\ket{L}+\ket{R})$. The second beam splitter makes detectors sensitive to state $\ket{\pm i}=1/\sqrt{2}(\ket{L}\pm i\ket{R})$ and we expect all three electrons to arrive at the same detector (regardless which one). If the electrons initially form a perfect triangle, we expect, as pointed out in Ref. \cite{pigeon}, the output and after the post-selection exhibits an enlarged triangle (due to all three particles in one arm) and disfigured individual spots (due to pairs propagating together). 

However, if we apply the weak value description, the Authors of the Ref. \cite{pigeon} conclude that no pair of particles has traveled through the same path. The projector certifying that electrons 1 and 2 travel through the same is given by
\begin{eqnarray}
\Pi^{same}_{12}=&&\Pi^{LL}_{12}+\Pi^{RR}_{12}\nonumber\\
=&&\left(\ket{LL}\bra{LL}+\ket{RR}\bra{RR}\right)\left(\ket{L}\bra{L}+\ket{R}\bra{R}\right).\nonumber\\
\end{eqnarray}
Importantly, this projector does not reveal which path was taken. The weak value is $\frac{\bra{+i+i+i}\Pi^{same}_{12}\ket{+++}}{\braket{+i+i+i}{+++}}=0$. The same argument applies to other pairs.

Nevertheless, if we study the weak value of the projection $\Pi^{same}_{123}=\ket{LLL}\bra{LLL}+\ket{RRR}\bra{RRR}$, it turns out to be -1/2. The weak value of $\Pi^{same}_{12}$ has four terms, which cancel pairwise and the total. Two of theme enter the weak value of $\Pi^{same}_{123}$, which raises the inconsistency of this weak value interpretation.

At this point, it is interesting to consider a Hamming distance between various components of the state. It tells us, how many particles need to move around to get to other state, So, for example, the the distance between states $\ket{LRR}$	and $\ket{RRL}$ is 2. Notice that when we consider the post-selection of all electrons leaving the interferometer in the same direction, pairs of components having odd Hamming distance interfere neutrally, i.e., the probability of observing all of the at the same side behind the second beam splitter is 1/8. If it is 2 modulo 4 -- destructively, and constructively for the Hamming distance equal to 0 mod 4. The weak value for particles 1 and 2 was found to be 0, since two pairs of contributions cancel each other. If we consider propagation of three electrons in the same arm, they would have Hamming distance of  3.

\section{Gaussian profile description}
We shall now discuss the optical description of the experiment. The Hilbert space is divided into two main parts: eight-dimensional Hilbert space $\{\ket{L},\ket{R}\}^{\otimes 3}$ and the spaces of complex functions of pairs of real parameters, describing transversal modes of all three particles. We actually need only the product of the amplitudes of each electron, so effectively our Hilbert space is of eight-dimensional complex functions of six real arguments. 

We adopt the following assumptions:
\begin{enumerate}
\item{Similarly to the original proposal, we consider three distinguishable particles. In other words, we do not perform the Hartree-Fock antisymmetrization. Still,  we consider electrons as it is necessary for them to repulse each other.}
\item{Initially the electrons have a very small separation, as compared to the width of the beam. They form a perfect triangle}
\item{The transverse profile of each electron is Gaussian and rotationally invariant. It doesn't change or get resized during the propagation, other than being displaced accordingly to the interaction.}
\item{The momentum kicks due to the repulsion is very small, thus strictly real 2-dimensional Gaussian functions describe the profiles after the  repulsion adequately. Alternatively, we could consider the displacement in the momentum space.}
\end{enumerate}
At first, these assumption may seem to difficult to fulfilled simultaneously. However, together they make the calculations feasible, serving for a guide for our intuition. We will discuss their relaxation after the main result is presented.

We assume that the electrons form a perfet triangle with vertices on lines of three vectors, $\vec{v}_1=(1,0)$, $\vec{v}_2=(-1,\sqrt{3})/2$, and $\vec{v}_3=(-1,-\sqrt{3})/2$, The beams have Gaussian profiles with spread $\sigma$. Thus, the state behind the first beam splitter is
\begin{equation}
\ket{\phi_0}=\frac{1}{2\sqrt{2}}\left(\begin{array}{c}1\\1\end{array}\right)^{\otimes 3}\times G,
\end{equation}
where $G=\prod_{i=1}^3F(x_i,y_i,(0,0))$ and $F(x,y,\vec{v})=\frac{1}{\sigma\sqrt{\pi}}\exp\left(-\frac{|(x,y)-\vec{v}|^2}{2\sigma}\right)$, and displacement due to traveling through different arms of the interferometer are already included in each of the components. After the repulsion and the second beam splitter we have
\begin{widetext}
\begin{equation}
\ket{\phi_1}=\frac{1}{2\sqrt{2}}\left(\begin{array}{c}F(x_1,y_1,a \vec{v}_1)F(x_2,y_2,a \vec{v}_2)F(x_3,y_3,a \vec{v}_3)\\
F(x_1,y_1,a (\vec{v}_1+\vec{v}_3)F(x_2,y_2,a(\vec{v}_2+\vec{v}_3))F(x_3,y_3,(0,0))\\
F(x_1,y_1,a(\vec{v}_1+\vec{v}_2)F(x_2,y_2,(0,0))F(x_3,y_3,a(\vec{v}_2+\vec{v}_3)\\
F(x_1,y_1,(0,0))F(x_2,y_2,a (\vec{v}_1+\vec{v}_2))F(x_3,y_3,a(\vec{v}_1+\vec{v}_3))\\
F(x_1,y_1,(0,0))F(x_2,y_2,a (\vec{v}_1+\vec{v}_2))F(x_3,y_3,a(\vec{v}_1+\vec{v}_3))\\
F(x_1,y_1,a(\vec{v}_1+\vec{v}_2)F(x_2,y_2,(0,0))F(x_3,y_3,a(\vec{v}_2+\vec{v}_3)\\
F(x_1,y_1,a (\vec{v}_1+\vec{v}_3)F(x_2,y_2,a(\vec{v}_2+\vec{v}_3))F(x_3,y_3,(0,0))\\
F(x_1,y_1,a \vec{v}_1)F(x_2,y_2,a \vec{v}_2)F(x_3,y_3,a \vec{v}_3)
\end{array}\right).
\end{equation}
\end{widetext}
Finally, we can compute the probability that all three electrons reach the same detector, $P_{LLL}$. 
\begin{eqnarray}
P_{LLL}=&&K(|(1,i,i,-1,i,-1,-1,-i)\ket{\phi_1}|^2)/8
\end{eqnarray}
with
\begin{eqnarray}
&&K(f(x_1,x_2,x_3,y_1,y_2,y_3))\nonumber\\
=&&\int_{-\infty}^{\infty}dx_1\int_{-\infty}^{\infty}dx_2\int_{-\infty}^{\infty}dx_3\nonumber\\
\times&&\int_{-\infty}^{\infty}dy_1\int_{-\infty}^{\infty}dy_2\int_{-\infty}^{\infty}dy_3\nonumber\\
\times&&f(x_1,x_2,x_3,y_1,y_2,y_3).
\end{eqnarray}
This probability is highly relevant, as the the quantum pigeonhole effect is conditioned on observing all three particles at the same time.

The amplitude in the modulo reads
\begin{eqnarray}
&&(1,i,i,-1,i,-1,-1,-i)\ket{\phi_1}\nonumber\\
=&&e^{-\frac{x_1^2+x_2^2+x_3^2+y_1^2+y_2^2+y_3^2+5a^2+a(-2x_1+x_2+x_3-\sqrt{3}y_2-\sqrt{3}y_3)}{\sigma^2}}\nonumber\\
\times&&(1+i)\left(\frac{1}{\sqrt{\pi}\sigma}\right)^3\left(-e^{\frac{a^2}{\sigma^2}}+e^{\frac{3a^2+a(x_1+x_2+x_3)}{2\sigma^2}}\right.\nonumber\\
+&&e^{\frac{a(x_1+x_2+x_3-\sqrt{3}y_1+\sqrt{3}y_2-\sqrt{3}y_3)}{\sigma^2}}\nonumber\\
+&&\left.e^{\frac{a(x_1+x_2+x_3-\sqrt{3}y_1-\sqrt{3}y_2+\sqrt{3}y_3)}{\sigma^2}}\right).\nonumber\\
\end{eqnarray}
After transformations we find that
\begin{equation}
P_{LLL}=\frac{1}{8}\left(1+\frac{3}{2}e^{-\frac{5}{8}x^2}-\frac{3}{2}e^{-\frac{3}{8}x^2}\right),
\end{equation}
where $x=a/\sigma$. This function reaches its minimum of $1-\left(\frac{3}{5}\right)^\frac{5}{2}$ for $x=2\sqrt{\log{5/3}}$.

Unfortunately, Ref. \cite{pigeon} does not offer a constructive explanation of the quantum pigeonhole effect. That is to say, we do learn from the paper how three electrons could have traveled without sharing a path. Also, within the above formalism, the projection onto not having more than one particle in each arm has no sense. Instead, we can verify a pair of particles have interacted. We have demonstrated above, that even in the weak value description, we can have three parties in one arm. We thus devise a test for exactly two particles propagating together. If such a pair travels together, it breaks the rotational symmetry of the output pattern. We thus define the probability of certifying the interaction as
\begin{eqnarray}
P_{INT}=1-\left|\bra{\phi_1}U\ket{\phi_1}\right|^2,
\end{eqnarray} and $U$ is composed of a rotation matrix by $2\Pi/3$ and a cyclic permutation $(x_1,y_1)\rightarrow(x_3,y_3)\rightarrow(x_2,y_2)\rightarrow(x_1,y_1)$. After plugging in the state we get
\begin{widetext}
\begin{eqnarray}
&&P_{INT}\nonumber\\
=&&1\nonumber\\
-&&\frac{1}{4}|K(G\times F(x_1,y_1,a (\vec{v}_1+\vec{v}_3))F(x_2,y_2,a(\vec{v}_2+\vec{v}_3))F(x_3,y_3,(0,0)))|^2\nonumber\\
-&&\frac{1}{4}|K(G\times F(x_1,y_1,a(\vec{v}_1+\vec{v}_2))F(x_2,y_2,(0,0))F(x_3,y_3,a(\vec{v}_2+\vec{v}_3)h))|^2\nonumber\\
-&&\frac{1}{4}|K(G\times F(x_1,y_1,(0,0))F(x_2,y_2,a(\vec{v}_1+\vec{v}_2))F(x_3,y_3,a(\vec{v}_1+\vec{v}_3)))|^2\nonumber\\
-&&\frac{1}{4}|K(G\times F(x_1,y_1,a\vec{v}_1)F(x_2,y_2,a\vec{v}_2)F(x_3,y_3,a\vec{v}_3))|^2\nonumber\\	
=&&\frac{3}{4}(1-e^{-\frac{5}{4}x^2}).
\end{eqnarray}
\end{widetext}

Note that $U$ acts on the profile degrees of freedom, whereas the beam splitters act solely on the discrete degrees. We have thus calculated just in front of the second beam splitter, rather than behind it. This also means that it  also commutes with the projectors $\Pi^{same}_{1,2}$ and $\Pi^{same}_{1,2,3}$.

When the interaction is strong, the state detection of even a single electron with a spatially-resolving detector reveals, which electrons shared a path, but not which path was more occupied. In other words, the state decoheres to two-dimensional block given by Hamming distance equal to 3. This lead to neutral decoherence, and equal probability for any output combinations in terms of $\ket{L}$ or $\ket{R}$ states behind the second beam splitter. For $x\rightarrow 0$ we do not observe symmetry braking, as the interaction is simply to weak, rather than being absent.

Let us now discuss possible relaxation of the assumptions. When we do not assume that the initial separation is equal to zero, we can assume that the initial wave-function is similar to the one with $x_0<x$. In such a case, some weak interaction has already taken place and we expect to get $\left.\partial\left(|\braket{\phi_i}{\phi_f}|^2\right)/\partial x\right|_{x=x_0}\neq 0$. Allowing for distortion of the spots and appearance of imaginary part due to repulsion can only decrease the overlaps. That means that all the effects are more dramatic: the dip of the probability for detecting all three particles is more prominent and the rise of probability of confirming the interaction grows more rapidly. However,  the asymptotic behavior remains unchanged.

We shall now discuss Refs. \cite{PNAS,EXP1} , which experimentally verifies the pigeonhole effect. Therein, the three particles were three photons propagating along three paths. Their polarizations are the analogue of the paths inside the interferometer in the original proposal. A single polarizing beam splitter (PBS) between any two represents the measurement of $\Pi^{same}_{i,j}$, while a sequence of two PBSs combining all three paths realizes $\Pi^{same}_{1,2,3}$. Under the condition that one photon reaches each detector and given the initial state, the mean of the former has been found to be 0, while the mean of the latter has been found to be about 0.23. This is another example of how this effect can be seen as a destructive interference, much like disappearing (dark) fringes when we close one slit in Young's experiment.

It should be stressed that the setup with one or two PBSs are two different strong measurements. Thus, in the Copenhagen interpretation the results are fully consistent, as they describe two distinct experiments.h

\section{Conclusions}
In conclusion, we have revisited the the experiment supposedly revealing the quantum pigeonhole effect by introducing the Gaussian transverse profiles of the electrons traveling through the Mach-Zehnder interferometer. We have first observed that the second statement of the effect, namely that the first derivative of all relevant probabilities with respect to the interaction strength vanishes, is a typical behavior in quantum mechanics. Second, there is no concept of an experiment to confirm the the claim that no two particles have traveled together. This remains an  interpretative conclusion of the two-vector state formalism. As this formalism allows projectors to yield negative values, it turned out that it is not-self consistent, giving the weak value of two electrons propagating together equal to zero (which is interpreted as an impossibility of at least two electrons traveling together), but the weak value	of the projection onto three electrons in one arm is different than zero. For weak interaction we do not see any effects, but this is due to the interaction having not enough time to change the state. For mid-range interaction strengths, we have a dip of probability of observing all three electrons at the same detector. For strong interaction, we observe the revival of three fold coincidences, as the repulsion causes a partial decoherence. Only the components with the Hamming distance of three will interfere.

The result presented above highlight a few things. First, weak values cannot be interpreted realistically, as this leads to inconsistencies, even within the weak value description itself. We have experienced it by finding a more general weak value  to be zero, but a more specific one - different. Even the fact that they can be arbitrary complex number suggests no direct operational meaning. Only the results of sharp, von Neumann measurements provide us the knowledge about the system's state, but one should be careful with the experimental context of such measurements. 

Second, it is logically incorrect to conclude that, since we have not observed the effects of interaction between particles, there wasn't any. This mistake was made, for example, Expression ``we have not observed'' is subjective and depends on our perceptive skills. The only way we could reach the conclusion of absence of the paairwise interaction is by designing a specific test for absence of such an interaction. Otherwise, it could be present, but simply too weak to bring in any observable effects. Thus far, claimed	experimental demonstrations did not include such a test. Interestingly, Ref. \cite{pigeon} mentions the deformation of the output beam as the method to verify the effect. The deformation would become clearer for longer interferometers. However, this test specifically requires electrons as pigeons.
\section*{Acknowledgements}
This work is a part of NCN grants 2017/26/E/ST2/01008 and 2015/19/B/ST2/01999, and joint DFG/NCN grant 
2016/23/G/ST2/04273. The Author acknowledges the ICTQT IRAP project of FNP (Contract No. 2018/MAB/5), financed by structural funds of EU.

\end{document}